\documentclass[sigconf,natbib=true]{acmart}
\AtBeginDocument{%
  }

\usepackage{graphicx}

\usepackage{amsmath}
\usepackage{xspace}
\usepackage{makecell}
\usepackage{adjustbox}
\usepackage{comment}
\usepackage{multirow}
\usepackage{caption}
\newcommand{\TempDPR}{\texttt{TempRetriever}\xspace}
\newcommand{\Vanilla}{\texttt{VanillaDPR}\xspace}
\newcommand{\DateToken}{\texttt{DateAsToken}\xspace}
\newcommand{\DateTag}{\texttt{DateAsTag}\xspace}
\setcopyright{acmlicensed}
\copyrightyear{2018}
\acmYear{2018}
\acmDOI{XXXXXXX.XXXXXXX}
\acmConference[Conference acronym 'XX]{Make sure to enter the correct
  conference title from your rights confirmation email}{June 03--05,
  2018}{Woodstock, NY}
\acmISBN{978-1-4503-XXXX-X/2018/06}

\begin{document}

\title{TempRetriever: Fusion-based Temporal Dense Passage Retrieval for Time-Sensitive Questions}


\author{Abdelrahman Abdallah}
\orcid{0000-0001-8747-4927}
\affiliation{%
  \institution{University of Innsbruck}
  \city{Innsbruck}
  \state{Tyrol}
  \country{Austria}
  }
\email{abdelrahman.abdallah@uibk.ac.at}

\author{Bhawna Piryani}
\orcid{0009-0005-3578-2393}
\affiliation{%
  \institution{University of Innsbruck}
  \city{Innsbruck}
  \state{Tyrol}
  \country{Austria}
  }
\email{bhawna.piryani@uibk.ac.at}

\author{Jonas Wallat}
\affiliation{%
  \institution{L3S Research Center}
 \city{Hannover }
  \country{Germany}
  }
\email{jonas.wallat@l3s.de}

\author{Avishek Anand}
\affiliation{%
  \institution{Delft University of Technology}
  \city{Delft}
  \country{The Netherlands}
  }
\email{	avishek.anand@tudelft.nl}

\author{Adam Jatowt}
\orcid{0000-0001-7235-0665}
\affiliation{%
  \institution{University of Innsbruck}
  \city{Innsbruck}
 \state{Tyrol}
  \country{Austria}
  }
\email{adam.jatowt@uibk.ac.at}

\renewcommand{\shortauthors}{Trovato et al.}

\begin{abstract}
Temporal awareness is crucial in many information retrieval tasks, particularly in scenarios where the relevance of documents depends on their alignment with the query's temporal context. Traditional approaches such as BM25 and Dense Passage Retrieval (DPR) focus on lexical or semantic similarity but tend to neglect the temporal alignment between queries and documents, which is essential for time-sensitive tasks like temporal question answering (TQA). We propose \TempDPR\footnote{The code and the datasets will be available after paper acceptance.}, a novel extension of DPR that explicitly incorporates temporal information by embedding both the query date and document timestamp into the retrieval process. This allows retrieving passages that are not only contextually relevant but also aligned with the temporal intent of queries. We evaluate \TempDPR on two large-scale datasets—ArchivalQA and ChroniclingAmericaQA—demonstrating its superiority over baseline retrieval models across multiple metrics. \TempDPR achieves a 6.63\% improvement in Top-1 retrieval accuracy and a 3.79\% improvement in NDCG@10 compared to the standard DPR on ArchivalQA. Similarly, for ChroniclingAmericaQA, \TempDPR exhibits a 9.56\% improvement in Top-1 retrieval accuracy and a 4.68\% improvement in NDCG@10.  We also propose a novel, time-based negative sampling strategy which further enhances retrieval performance by addressing temporal misalignment during training. Our results underline the importance of temporal aspects in dense retrieval systems and establish a new benchmark for time-aware passage retrieval.
\end{abstract}




\maketitle

\section{Introduction}
\label{sec:intro}

Time is a crucial dimension in nearly any information space and plays a significant role in document and language understanding~\cite{dhingra2022time,tan2023towards,yuan2023zero,saxena2021question,jia2021complex}. Its correct identification and use are fundamental to various information retrieval and natural language processing systems and applications~\cite{chen2021dataset,mavromatis2022tempoqr,pasca2008towards,wang2020answering,yuan2024back,abdallah2023exploring} such as ones operating over news articles, historical records, biographies, research papers, legal documents, etc. 
As Metzger~\cite{metzger2007making} noted, timeliness is a significant measure of the quality of information, alongside accuracy, coverage, objectivity, and relevance. Therefore, incorporating temporal signals into the retrieval process is essential to improve the quality of retrieved results in time-sensitive contexts. For example, consider the queries \textit{What are the key policies of the current US President?} versus \textit{What were the key policies implemented by the US President in 2022?} The first query seeks an overview of the current President’s policies, potentially covering the entire scope of the presidency. In contrast, the second query focuses on policies explicitly implemented in 2022, requiring the retrieval system to prioritize relevant documents from that particular time.

 
Recent retrieval methods such as ones using dense retrievers like Dense Passage Retrieval (DPR)~\cite{karpukhin2020dense,ren2021rocketqav2,tang2022dptdr}
lack however explicit temporal information encoding. This limitation hinders their effectiveness in handling time-sensitive queries. As a result, while these systems excel in open-domain question answering (QA), they do not recognize and capture the importance of time in \textit{temporal question answering} (TQA) and other temporal IR scenarios. 

In this paper, we explore the following question: \textit{Can we develop an effective dense retrieval model that incorporates temporal information alongside the semantic representation of queries and passages?} 
We introduce three novel approaches for integrating temporal aspects into the DPR framework. First, the \textbf{\DateTag} approach wraps timestamps in special tokens, such as \texttt{[S-DATE]} and \texttt{[E-DATE]}, to explicitly mark temporal information in the input. Second, in the \textbf{\DateToken} approach, we append the timestamp (e.g. year, month, or full date) as plain text to the query and document, treating it as part of the input text.  Third, our core contribution, \textbf{Temporal Dense Passage Retrieval (\TempDPR)}, incorporates temporal embeddings directly into the query and document representations. To achieve this, we explore various fusion techniques, including Feature Stacking (FS), Vector Summation (VS), Relative Embeddings (RE), and Element-Wise Interaction (EWI). These techniques enable the model to encode temporal information within dense embeddings for ensuring robust temporal alignment. 

In this work, we mainly focus on questions containing explicit dates (referred to explicit temporal questions) as they are relatively common in TQA and are easier to handle. However, to emphasize the generalizability of our proposal we also demonstrate how to extend it to the case where questions do not contain any explicit temporal expressions (i.e., implicit temporal questions). 
%
%
%
%
\begin{figure*}[!t]
\centering
\includegraphics[width=0.7\textwidth]{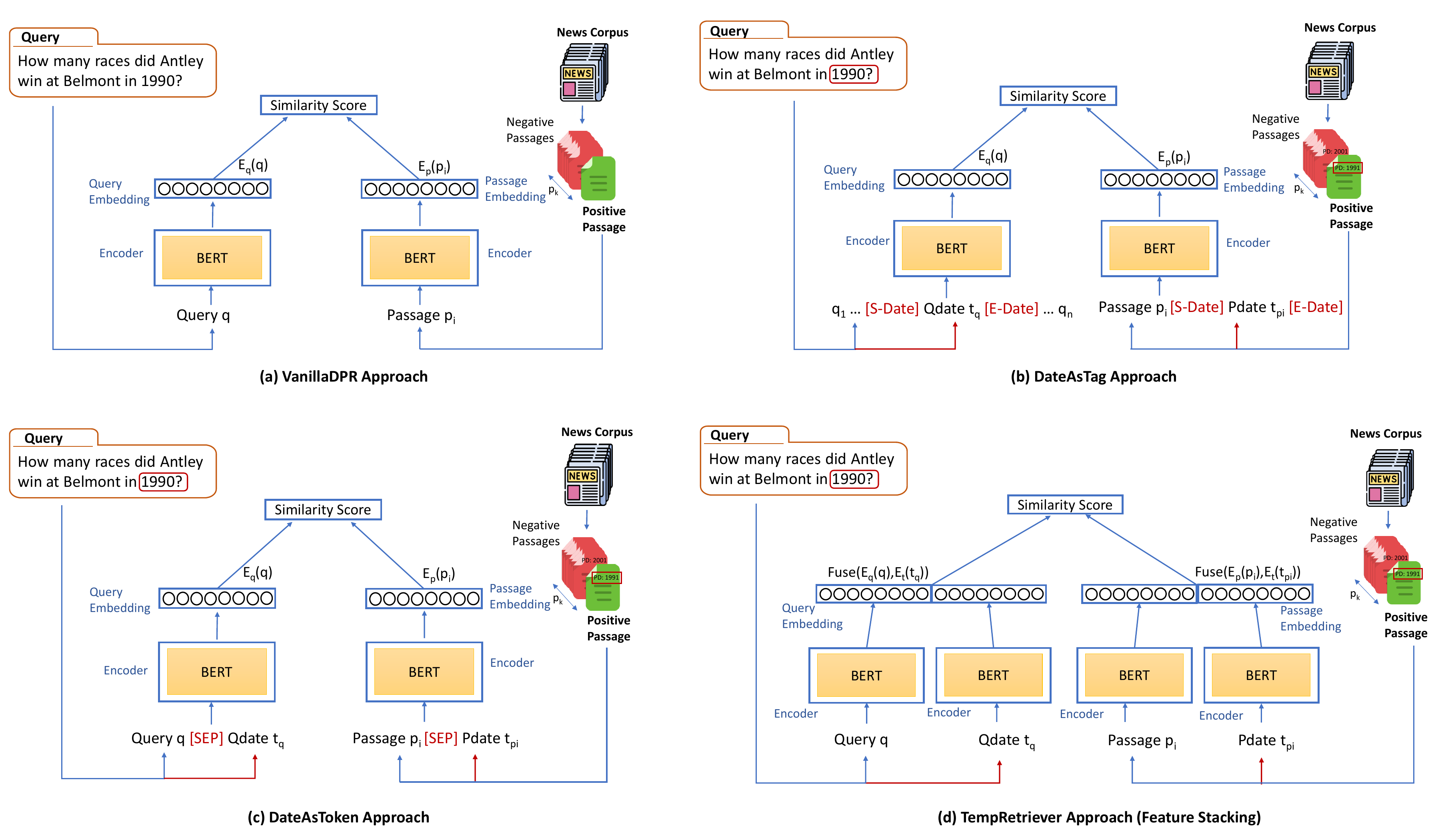}
\caption{A schematic diagram illustrating an overview of training used in different approaches: (a) \Vanilla approach, 
(b) \DateTag approach, 
(c) \DateToken approach, 
and (d) \TempDPR. 
Due to space limitation we only show here the feature stacking technique for the \TempDPR approach. 
Note that the positive passage's publication date (PD) may differ from the date in the query, as demonstrated also in the figure's example.}
\label{fig:Approaches}
\end{figure*}

To sum up, our contributions are fourfold: 
\begin{enumerate}
    \item First, we propose incorporating the \textbf{question date} and \textbf{passage date embedding} into the existing question-passage framework. We then demonstrate that the \textbf{proposed approach} significantly outperforms standard DPR on two TQA datasets. 
    \item Second, we propose a \textbf{time-based hard negative sampling strategy} which selects negative passages based on their temporal characteristics. We experiment with three different approaches: (1) randomly selecting hard negatives, (2) selecting negative documents from the same year as the positive document, and (3) selecting documents from a year different from the positive document's year. 
    \item Third, we integrate \TempDPR into a \textbf{Retrieval-Augmented Generation (RAG)} pipeline. Using large language models (LLMs), we explore the effectiveness of \TempDPR's retrieved results in improving answer generation. 
    \item Finally, we demonstrate how our approach can be generalized to be used for \textbf{questions lacking explicit temporal information}.

\end{enumerate}

\section{Related Work}
The research on \textit{Temporal Information Retrieval }(TIR)~\cite{campos2015survey,kanhabua2016temporal,omar_spaniol2024,jatowt2005temporal} emerged in the past as a specialized type of information retrieval, focusing on searching documents with respect to time dimension and temporal signals. While the previous research primarily addressed temporal representation and understanding of queries and documents, there has been limited exploration of integrating temporal information with modern neural network-based retrieval technologies and its application in the context of question answering. Our work seeks to fill that gap by incorporating temporal embeddings and alignment into dense retrieval architectures.

Passage retrieval is a critical component in open-domain question answering, as the efficiency and precision of retrieving relevant passages directly impact the overall performance of the system \cite{chen-etal-2017-reading,chen-yih-2020-open,jia2018tempquestions,dhingra2022time}. Both sparse and dense vector models have been developed to enhance passage retrieval. Sparse retrieval methods represent queries and documents as sparse vectors, where each element corresponds to a term in the vocabulary. Dense retrieval methods, on the other hand, encode queries and documents into a continuous vector space, often leveraging pre-trained language models to capture richer semantic representations.

Notable contributions to dense passage retrieval (DPR) include various architectures designed to improve the representation and retrieval of passages in QA tasks \cite{karpukhin2020dense, yang-etal-2019-end-end, min-etal-2019-discrete, wolfson-etal-2020-break,jia2018tequila,shang2022improving}. These models utilize deep learning techniques to encode queries and documents as dense vectors, enabling more effective semantic matching compared to traditional lexical methods. Despite these advancements, dense retrieval models often overlook temporal signals, a key aspect of many time-sensitive information retrieval tasks.

In the recent years, several studies~\cite{xiong2024large,yuan2024back,tan2023towards,chu2023timebench,su2024timo,wei2023menatqa} have explored the integration of temporal reasoning into language models to address the challenges posed by dynamic information needs. 
Cole et al. ~\cite{cole-etal-2023-salient} proposed a novel pre-training technique called Temporal Span Masking (TSM), which enhances temporal reasoning by masking and predicting temporal spans such as dates and periods during training. This approach improves the model's ability to understand and reason about temporal relationships.
Other works have focused on adapting language models to handle time-sensitive information. For example, Giulianelli et al. \cite{giulianelli-etal-2020-analysing} analyzed contextualized word embeddings of BERT to track lexical semantic change over time, demonstrating how word meanings shift across different time periods. Rosin and Radinsky \cite{rosin-radinsky-2022-temporal} introduced temporal attention mechanisms, explicitly incorporating temporal signals into the attention computation. Building on this idea, Rosin~\cite{rosin-2022} further proposed methods to enhance temporal reasoning by masking time-related tokens during training. Wang et al. \cite{wang2023bitimebert} introduced BiTimeBERT a model that integrates document timestamps and temporal references within text based on time-aware masked language modeling and document dating pretraining tasks. This dual temporal approach significantly improved tasks like event time prediction and temporal question answering by enhancing the model’s ability to understand how events evolve and relate to various temporal contexts.  Wu et al. \cite{wu2024time} proposed a framework, called TS-Retriever, for time-sensitive retrieval tasks. The proposed method integrates supervised contrastive learning with tailored positive and negative sample pairs, emphasizing temporal constraints. Unlike Karpukhin et al. \cite{karpukhin2020dense} they alter the time specifiers in the queries to generate the mismatched negative pairs.

While previous studies demonstrate the potential of temporal signals in improving reasoning capabilities of  language models, their application to dense passage retrieval remains limited. Existing dense retrieval models, such as DPR, have yet to fully exploit temporal information to enhance retrieval performance for time-sensitive tasks. To bridge this gap, our work introduces a novel temporal passage retrieval framework that leverages temporal signals to provide more accurate and contextually relevant retrieval for QA systems.

\section{Method}



We first experiment with two simple mechanisms to incorporate temporal information into the retrieval process. 

In the first one, referred to as the \DateTag approach and shown in Figure \ref{fig:Approaches}b, we explicitly mark dates in the input question using special tokens, formatted as \texttt{q$_1$ ... [S-DATE] Qdate [E-DATE] ... q$_n$ } where \texttt{q$_1$} and \texttt{q$_n$} are the first and last token of the question, respectively, and \texttt{Qdate}\footnote{In this work, we consider queries that contain only one explicit date mention.} is the date mentioned in the query. For a passage, we append \texttt{[S-DATE] Pdate [E-DATE]} where \texttt{Pdate} is the timestamp of the document containing the passage (publication date). The second mechanism, the \DateToken approach (Figure \ref{fig:Approaches}c), appends the dates to the query and passage as plain text, formatted as \texttt{query [SEP] Qdate} and as \texttt{passage [SEP] Pdate}.

Building on these, we next propose a \textbf{Temporal Dense Passage Retrieval (\TempDPR)}, which explicitly incorporates both the content and temporal information of queries and passages. Unlike the \DateTag and \DateToken approaches, \TempDPR represents temporal information as continuous embeddings, enabling it to effectively capture temporal relationships in a learned vector space. To integrate semantic and temporal embeddings, we introduce several fusion techniques, such as feature stacking, vector summation, relative embeddings, and element-wise interaction. These fusion operations allow the model to combine semantic and temporal features in different ways during the retrieval process.
All the approaches tested in this paper, including the proposed \TempDPR , are illustrated in Figure \ref{fig:Approaches}. 


\subsection{Temporal Dense Passage 
Retrieval}
\label{Temporal Dense Passage Retriever}

Let $\mathcal{V}$ represent the vocabulary space encompassing all possible tokens within a corpus. We define $\mathcal{P} = \{p_1, p_2, \dots, p_M\}$ be a collection of $M$ text passages, where each passage $p \in \mathcal{P}$ is a sequence of tokens drawn from $\mathcal{V}$. Similarly, $\mathcal{Q}$ denotes the set of all possible queries with each query $q \in \mathcal{Q}$ also being a sequence of tokens from $\mathcal{V}$. We further introduce the temporal space $\mathcal{T} = \{t_{p1}, t_{p2}, \dots, t_{pM}\}$ as the corresponding set of timestamps associated with each passage. 
Our goal is to encode both the text passage $p_i$ and its timestamp $t_{pi}$ into a low-dimensional continuous space, $\mathbb{R}^d$, that captures the complex interplay of semantic and temporal nuances. 

For each passage $p_i$ and its timestamp $t_{pi}$, we define the encoding function $f_p(p_i, t_{pi})$:
\[
f_p(p_i, t_i) = \mathrm{Fuse}(E_p(p_i), E_t(t_{pi})),
\]
where $E_p(\cdot)$ encodes the passage content, $E_t(\cdot)$ encodes the timestamp, and $\mathrm{Fuse}(\cdot)$ represents the fusion operation used to combine the two embeddings. To jointly model semantic and temporal information, we define three encoding functions: a passage encoder, a query encoder, and a temporal encoder. 

The encoding framework in \TempDPR consists of components designed to represent both the textual and temporal aspects of queries and passages. A single pretrained BERT model is utilized, producing semantic embeddings for both queries and passages. Specifically, the passage encoder $E_p$ maps a passage $p \in \mathcal{P}$ into a semantic embedding space $\mathbb{R}^{d_s}$, and the same model is used to map a query $q \in \mathcal{Q}$ into the same embedding space $\mathbb{R}^{d_s}$. The final representation is derived from the \texttt{[CLS]} token, where $E_p(p) = \text{BERT}(p)_{\texttt{[CLS]}}$ and $E_q(q) = \text{BERT}(q)_{\texttt{[CLS]}}$. 
In addition to semantic embeddings, the temporal encoder $E_t$ maps timestamps $t \in \mathcal{T}$ into a temporal embedding space $\mathbb{R}^{d_t}$, allowing temporal information to complement the semantic representations. 
The semantic and temporal embeddings are then integrated using one of four fusion operations to create a unified representation for downstream retrieval tasks.


(1) \textbf{Vector Summation} (\textbf{VS}): Combines the embeddings through element-wise addition:
    \[
    \mathrm{Fuse}(x, y) = x + y.
    \]

(2) \textbf{Relative Embeddings} (\textbf{RE}): Captures the relative relationship between embeddings using element-wise subtraction:
    \[
    \mathrm{Fuse}(x, y) = x - y.
    \]

(3) \textbf{Element-Wise Interaction} (\textbf{EWI}): Highlights feature interactions by performing element-wise multiplication:
    \[
    \mathrm{Fuse}(x, y) = x \cdot y.
    \]

(4) \textbf{Feature Stacking} (\textbf{FS}): Combines the two embeddings by appending them along the feature dimension:
    \[
    \mathrm{Fuse}(x, y) = [x \oplus y],
    \]
Similarly, for a query $q$ and its timestamp $t_q$, we define the query encoding as:
\[
f_q(q, t_q) = \mathrm{Fuse}(E_q(q), E_t(t_q)),
\]
where $E_q(\cdot)$ encodes the query content and $E_t(\cdot)$ encodes the query date. 
These fusion techniques allow us to explore different ways of combining temporal and semantic information, ensuring that both are effectively represented in the final embedding. 

For \TempDPR, we aim to optimize the balance between representing semantic relevance and the ability to capture temporal relevance. Temporal information adds complexity to the retrieval task, as both the content and the time dimension between the query and its matching passages must be aligned. A variety of similarity metrics can be employed for this purpose. We focus on the inner product of the fused semantic and temporal vectors for its simplicity and efficiency.
%
The relevance score between a query $q$ and a passage $p_i$ is then computed using the dot product of their encoded vectors:
\[
\mathrm{sim}(q, p_i) = f_q(q, t_q)^\top f_d(p_i, t_{pi}),
\]
where $f_q(q, t_q)$ and $f_d(p_i, t_{pi})$ represent the fused query and document embeddings, respectively.

In \TempDPR, the temporal alignment is achieved by embedding timestamps into a shared representation space. Specifically, we encode the query timestamp ($t_q$) and passage timestamp ($t_{pi}$) alongside their respective text embeddings, ensuring that temporal signals are weighted equally with semantic content during the similarity computation. We also introduce in Sec. \ref{sec:negativepassages} temporal negative sampling strategy to address the common issue of temporal misalignment in retrieval tasks, explicitly training the model to distinguish between temporally relevant and irrelevant documents. Consequently, \TempDPR can prioritize documents with temporal alignment while maintaining textual relevance. We also note here that the publication date (PD) of a relevant passage does not necessarily need to be the same as the date mentioned in the question. The information relevant to an event is often published around the time of the event, and hence the PD of a positive passage may not always align exactly with the date referenced in the question. 


\subsection{Training}
In \TempDPR model, we employ a BERT-based encoder~\cite{devlin2019bertpretrainingdeepbidirectional} to capture both the semantic content of the query and passage, as well as their temporal contexts. Unlike conventional retrieval models that rely purely on semantic embeddings, our approach extends the representation by incorporating temporal information using various fusion techniques. 

For feature stacking, the combined representation has a latent space dimension of \(d = {d_s} + {d_t} = 1,536\), as the semantic and temporal embeddings are appended along the feature dimension. For other fusion techniques, such as vector summation, relative embeddings, and element-wise interaction, the representation remains in a latent space of \(d = 768\), as the fusion operation combines the embeddings without increasing the dimensionality.


To learn the parameters of the \TempDPR retrieval model, we formulate a training objective that explicitly considers both semantic and temporal relevance. The training dataset $\mathcal{D} $, defined as $\mathcal{D} = \{ \langle q_i, t_{q_i}, p^+_i, t_{p^+_i}, p^-_{i,1}, t_{p^-_{i,1}}, \dots, p^-_{i,n}, t_{p^-_{i,n}} \rangle \}_{i=1}^m$, where each instance in the dataset contains a query $q_i$ and its associated timestamp $t_{q_i}$, a positive passage $p^+_i$ with its timestamp $t_{p^+_i}$, and $n$ negative passages $\{p^-_{i,j}, t_{p^-_{i,j}}\}$ each with their own timestamps. Our goal is to optimize the retrieval such that the positive passage, which is temporally aligned with the query, is ranked higher than the negative passages. To train the model, we use a cross-entropy loss function that encourages the similarity score between the query and the positive passage to be higher than that between the query and the negative passages while accounting for temporal alignment. Formally, the loss for a single instance is defined as:

\noindent
\scalebox{0.8}{ 
\begin{minipage}{\columnwidth}
\begin{align*} 
L(q_i, p^+_i, p^-_{i,1}, \dots, p^-_{i,n}) 
&= - \log \frac{
    e^{\mathrm{sim}(f_q(q_i, t_{q_i}), f_d(p^+_i, t_{p^+_i}))}
}{
    e^{\mathrm{sim}(f_q(q_i, t_{q_i}), f_d(p^+_i, t_{p^+_i}))} 
    + \sum_{j=1}^n e^{\mathrm{sim}(f_q(q_i, t_{q_i}), f_d(p^-_{i,j}, t_{p^-_{i,j}}))}
}.
\end{align*}
\end{minipage}
}


\section{Experimental Setup}

\subsection{Temporal Question Answering Datasets}

For our experiments, we use two QA datasets specifically curated for temporal IR tasks: ArchivalQA and ChroniclingAmericaQA. 

ArchivalQA~\cite{wang2022archivalqa} is a benchmark dataset designed for open-domain question answering over historical news collections. It comprises 
question-answer pairs with questions derived from a diverse set of news articles 
published between 1987 and 2007. The dataset is particularly suited for exploring the temporal dynamics of news data and for training models on short-term, historically grounded document collections. 

ChroniclingAmericaQA~\cite{piryani2024chroniclingamericaqa} is another large-scale dataset focused on question-answering over American historical newspaper collections. The dataset spans a longer period, from 1800 to 1920. 
The questions reflect the evolving historical and cultural context of the time, making the dataset highly valuable for tasks involving historical and archival documents. 
\begin{table}[h!]
    \scriptsize 
    \setlength\tabcolsep{5pt}
    \centering
    \caption{Dataset statistics for ArchivalQA and ChroniclingAmericaQA.}
    \resizebox{0.9\columnwidth}{!}{ 
    \begin{tabular}{llllll} \toprule
    Dataset & Type & Train & Val & Test \\
    \midrule
    ArchivalQA                  & Explicit      & 62,157    & 7,841     & 7,783\\
    ArchivalQA                  & Implicit          & 5,000   & 1,000    & 1,000   \\
    ChroniclingAmericaQA        & Explicit      & 22,918    & 1,232     & 1,268 \\
    \bottomrule
    \end{tabular}
    } 
    \label{tab:dataset-split}
\end{table}

Table \ref{tab:dataset-split} provides the dataset splits for both ArchivalQA and ChroniclingAmericaQA, detailing their training, validation, and testing sets following ~\cite{wallat2024temporal}. The datasets are divided into two subsets: Explicit and Implicit. The Explicit subset focuses on questions that explicitly reference temporal information, such as dates or temporal expressions. This subset is designed to evaluate the ability of retrieval models to handle time-sensitive queries directly. The Implicit subset consists of questions where the query does not explicitly mention any temporal information. To handle these questions, we developed a query routing and date prediction model, discussed in Sec. \ref{sec: implicit}, that predicts the relevant timestamp associated with the query~\cite{DBLP:conf/sigir/WangJY21,gupta2014identifying,das2017estimating}. This additional step enables temporal reasoning even for queries lacking direct temporal references, allowing the retrieval model to align the query with temporally relevant passages. The above approach ensures that the retrieval system can effectively generalize over explicit and implicit temporal queries.


\begin{table*}[ht!]
\centering
\caption{Top-\{1,5,10,20,50,100\} retrieval accuracy for \textit{ArchivalQA} and \textit{ChroniclingAmericaQA} datasets. Results are evaluated on both validation and test sets. \TempDPR variants consistently outperform state-of-the-art temporal retrieval models.}
\setlength\tabcolsep{5pt}
\resizebox{0.7\textwidth}{!}{  
\begin{tabular}{l|l|cccccc|cccccc} 
\hline
\multirow{2}{*}{\textbf{Dataset}} & \multirow{2}{*}{\textbf{Model}} & \multicolumn{6}{c|}{\textbf{Validation }} & \multicolumn{6}{c}{\textbf{Test }} \\ 
 
\cline{3-14}
 &  & Top-1 & Top-5 & Top-10 & Top-20 & Top-50 & Top-100 & Top-1 & Top-5 & Top-10 & Top-20 & Top-50 & Top-100 \\ 
\midrule
\multirow{7}{*}{\textbf{ChroniclingQA}} 
 & BiTimeBert~\cite{wang2023bitimebert} & 35.74 & 54.02 & 61.09   & 66.94 & 74.90 &78.31  &38.52  & 54.62 & 60.46   & 65.59 & 72.69  &76.95  \\

 & TS-Retriever~\cite{wu2024time} & 47.85 & 63.93 & 69.29  & 73.60  & 79.53 & 83.43  &46.65  & 62.35 & 69.06 &74.27  &78.77  &82.56  \\

& \Vanilla~\cite{karpukhin2020dense} & 44.97 & 61.14 & 67.86 & 73.32 & 79.12 & 83.53 & 45.70 & 60.47 & 67.40 & 72.20 & 78.69 & 82.26 \\ 
\cline{2-14}

 & \TempDPR$_{VS}$ & 49.23 & 64.74 & 70.51 & 75.14 & 81.80 & 85.38 & 49.72 & 65.11 & 71.27 & 76.72 & 81.45 & 84.77 \\
 & \TempDPR$_{RE}$ & 48.90 & 64.34 & 69.46 & 75.55 & 81.48 & 84.40 & 47.28 & 65.67 & 71.74 & 75.85 & 82.24 & 85.56 \\
 & \TempDPR$_{EWI}$ & 47.44 & 62.88 & 69.70 & 74.82 & 80.18 & 84.48 & 46.57 & 63.38 & 69.69 & 75.22 & 80.66 & 84.06 \\ 
& \TempDPR$_{FS}$ & \textbf{50.22 }& \textbf{64.88} & \textbf{71.16} & \textbf{77.90} & \textbf{83.05} & \textbf{86.22} & \textbf{50.10} & \textbf{66.01} & \textbf{71.85} & \textbf{77.19} & \textbf{82.81} & \textbf{86.52} \\ 
\midrule

\multirow{7}{*}{\textbf{ArchivalQA}} 
 & BiTimeBert~\cite{wang2023bitimebert} & 61.71 & 81.20 & 86.56 & 89.99 & 93.88 & 95.52 & 60.22 & 81.20 & 86.70 & 90.48 & 94.09 & 95.62  \\

 & TS-Retriever~\cite{wu2024time}  & 63.44 & 83.74 & 88.70 & 92.53 & 95.88 & 97.33 & 64.68 & 84.75 & 89.44 & 92.92 & 95.76 & 97.29 \\

 & \Vanilla~\cite{karpukhin2020dense} & 63.09 & 82.42 & 87.47 & 90.97 & 94.40 & 96.14& 62.98 & 82.50 & 87.17 & 90.97 & 94.64 & 96.15 \\ 

 \cline{2-14}

 & \TempDPR$_{VS}$ & 68.42 & 86.23 & 90.55 & 93.30 & 95.97 & 97.30 & 67.96 & 86.23 & 90.47 & 93.55 & 96.24 & 97.55 \\
 
 & \TempDPR$_{RE}$ & 66.76 & 86.93 & 90.98 &94.02  & 96.40 &  97.46 &  66.11 & 87.13 & 91.53 & 94.32 &  96.83 &  97.88 \\

 & \TempDPR$_{EWI}$ & \textbf{69.97} & 85.40 &  90.18 & 93.34 & 95.71 &  97.08 & 69.27 & 85.79 & 90.04 & 93.05 & 95.85 &  97.16 \\

 & \TempDPR$_{FS}$ & 69.72 & \textbf{87.82} & \textbf{91.38} &  \textbf{94.21} & \textbf{96.44} & \textbf{97.77} &  \textbf{69.84} & \textbf{87.78} & \textbf{91.65} & \textbf{94.28 }& \textbf{96.78} & \textbf{97.75}\\
 
\bottomrule
\end{tabular}
 }

\label{tab:combined_results}
\end{table*}

\begin{table*}[ht]
\centering
\caption{Top-\{1,5,10,20,50,100\} retrieval accuracy for \textit{ArchivalQA} and \textit{ChroniclingAmericaQA} datasets, evaluated on both validation and test sets. The results measure the percentage of top retrieved passages that contain the correct answer across different temporal retrieval models, including the baseline \Vanilla, \DateTag, \DateToken,  and \TempDPR approaches.}
\setlength\tabcolsep{5pt}
\resizebox{0.9\textwidth}{!}{  
\begin{tabular}{l|l|cccccc|cccccc} 
\hline
\multirow{2}{*}{\textbf{Dataset}} & \multirow{2}{*}{\textbf{Mode}} &  \multicolumn{6}{c|}{\textbf{Validation Results}} & \multicolumn{6}{c}{\textbf{Test Results}} \\ 
\cline{3-14}

\cline{3-14}
 &  & Top-1 & Top-5 & Top-10 & Top-20 & Top-50 & Top-100 & Top-1 & Top-5 & Top-10 & Top-20 & Top-50 & Top-100 \\ 
\midrule
\multirow{4}{*}{\textbf{ArchivalQA}} 
 & \Vanilla & $63.09_{\pm 0.76}$ & $82.42_{\pm 0.29}$ & $87.47_{\pm 0.11}$ & $90.97_{\pm 0.15}$ & $94.40_{\pm 0.21}$ & $96.14_{\pm 0.03}$ & $62.98_{ \pm 0.55}$ & $82.50_{ \pm 0.33}$ & $87.17_{ \pm 0.45}$ & $90.97_{ \pm 0.27}$ & $94.64_{ \pm 0.23}$ & $96.15_{ \pm 0.16}$ \\ 
 & \DateTag & $57.94_{\pm 0.71}$ & $78.78_{\pm 0.61}$ & $84.28_{\pm 0.41}$ & $88.48_{\pm 0.48}$ & $92.43_{\pm 0.26}$ & $94.64_{\pm 0.13}$ & $58.37_{ \pm 0.70}$ & $78.43_{ \pm 0.84}$ & $83.93_{ \pm 0.45}$ & $88.21_{ \pm 0.41}$ & $92.77_{ \pm 0.25}$ & $94.93_{ \pm 0.10}$ \\
 & \DateToken & $63.51_{\pm 0.32}$ & $83.41_{\pm 0.28}$ & $88.52_{\pm 0.08}$ & $91.97_{\pm 0.08}$ & $94.92_{\pm 0.05}$ & $96.52_{\pm 0.02}$ & $63.44_{ \pm 0.29}$ & $83.23_{ \pm 0.05}$ & $88.35_{ \pm 0.10}$ & $91.72_{ \pm 0.03}$ & $94.94_{ \pm 0.23}$ & $96.72_{ \pm 0.08}$ \\ 
  
 & \TempDPR$_{FS}$ & $\mathbf{69.72}_{\pm 0.34}$ & $\mathbf{87.82}_{\pm 0.32}$ & $\mathbf{91.38}_{\pm 0.09}$ & $\mathbf{94.21}_{ \pm 0.06}$ & $\mathbf{96.44}_{\pm 0.12}$ & $\mathbf{97.77}_{\pm 0.03}$ & $\mathbf{69.84}_{ \pm 0.86}$ & $\mathbf{87.78}_{ \pm 0.46}$ & $\mathbf{91.65}_{ \pm 0.36}$ & $\mathbf{94.28}_{ \pm 0.35}$ & $\mathbf{96.78} _{\pm 0.23}$ & $\mathbf{97.75}_{ \pm 0.13}$ \\

\midrule
\multirow{4}{*}{\textbf{ChroniclingQA}} 
 & \Vanilla & $44.97_{\pm 0.24}$ & $61.14_{\pm 0.16}$ & $67.86_{\pm 0.16}$ & $73.32_{\pm 0.21}$ & $79.12_{\pm 0.10}$ & $83.53_{\pm 0.15}$ & $45.70_{ \pm 0.46}$ & $60.47_{ \pm 0.25}$ & $67.40_{ \pm 0.14}$ & $72.20_{ \pm 0.18}$ & $78.69_{ \pm 0.13}$ & $82.26_{ \pm 0.09}$ \\ 
 & \DateTag & $45.00_{\pm 0.17}$ & $61.48_{\pm 0.31}$ & $67.62_{\pm 0.19}$ & $72.97_{ \pm 0.07}$ & $79.14_{ \pm 0.18}$ & $82.99_{ \pm 0.18}$ & $45.75_{ \pm 0.48}$ & $60.77_{ \pm 0.49}$ & $66.77_{ \pm 0.22}$ & $72.44_{ \pm 0.32}$ & $78.72_{ \pm 0.18}$ & $81.89_{ \pm 0.21}$ \\ 
 & \DateToken & $44.81_{\pm 0.04}$ & $61.28_{\pm 0.28}$ & $67.62_{\pm 0.16}$ & $73.60_{\pm 0.13}$ & $79.12_{\pm 0.25}$ & $83.17_{\pm 0.07}$ & $45.81_{ \pm 0.09}$ & $60.79_{ \pm 0.19}$ & $66.58_{ \pm 0.20}$ & $72.09_{ \pm 0.15}$ & $78.82_{ \pm 0.07}$ & $82.10_{ \pm 0.22}$ \\ 

 & \TempDPR$_{FS}$ & $\mathbf{50.22}_{\pm 0.19}$ & $\mathbf{64.88}_{\pm 0.23}$ & $\mathbf{71.16}_{\pm 0.30}$ & $\mathbf{77.90}_{ \pm 0.21}$ & $\mathbf{83.05}_{\pm 0.11}$ & $\mathbf{86.22}_{\pm 0.29}$ & $\mathbf{50.10}_{ \pm 0.10}$ & $\mathbf{66.01}_{ \pm 0.17}$ & $\mathbf{71.85}_{ \pm 0.39}$ & $\mathbf{77.19}_{ \pm 0.11}$ & $\mathbf{82.81}_{ \pm 0.25}$ & $\mathbf{86.52}_{ \pm 0.10}$ \\ 
\bottomrule
\end{tabular}
}

\label{tab:top-explicit_dataset}
\end{table*}

\subsection{Corpus Pre-processing}


The ArchivalQA and ChroniclingAmericaQA datasets are built on two distinct corpora. ArchivalQA is curated from the New York Times Annotated Corpus (NYT corpus)~\cite{sandhaus2008new}, which contains news articles spanning 1978 to 2007. On the other hand, ChroniclingAmericaQA is associated with its own corpus, Chronicling America \footnote{\url{https://chroniclingamerica.loc.gov/about/}}, which includes historical news content from 1800 to 1920.
To prepare the corpora for retrieval, we follow a standard preprocessing approach~\cite{karpukhin2020dense,wang2019multi}. Following the methodology used in DPR~\cite{karpukhin2020dense}, each article is split into multiple, disjoint text blocks of 100 words, with each block serving as a passage—our basic retrieval unit. This process yields a total of 19,851,114 passages for the NYT corpus.
In contrast, the ChroniclingAmericaQA corpus does not require the creation of passages, as it is already structured into smaller, self-contained passages each in the range of 250 words with a total of 151,485 passages for the ChroniclingAmericaQA corpus. To provide additional context, each passage is prepended with the article's title and separated by a [SEP] token.

\subsection{Selection of Positive Passages}

For ArchivalQA, the associated NYT corpus contains documents of varying lengths, requiring a passage-based approach for positive passage selection. 
A passage is positive if it contains the answer to the question. If the answer appears in multiple distinct passages within a document, all such passages are considered positives.
For ChroniclingAmericaQA dataset, not every question is guaranteed to have an answer within the provided candidate passages. To address this, we apply a filtering step to remove questions without corresponding answers in the candidate passages. This filtering ensures that both the training and testing sets include only questions with valid, retrievable answers, thereby maintaining the integrity of the evaluation process.

\begin{table*}[ht]
\centering
\caption{Results for nDCG@\{1,5,10,20,50,100\} and MAP@\{1,5,10,20,50,100\} for the \textit{ArchivalQA} and \textit{ChroniclingAmericaQA} datasets, across validation and test sets.}
\setlength\tabcolsep{5pt}
\resizebox{0.9\textwidth}{!}{  
\begin{tabular}{l|l|cccccc|cccccc|cccccc|cccccc} 
\hline
\multirow{3}{*}{\textbf{Retriever}} & \multirow{3}{*}{\textbf{Mode}} 
& \multicolumn{12}{c|}{\textbf{nDCG}} 
& \multicolumn{12}{c}{\textbf{MAP}} \\ 
\cline{3-26}
 &  & \multicolumn{6}{c|}{\textbf{Validation}} & \multicolumn{6}{c|}{\textbf{Test}} 
 & \multicolumn{6}{c|}{\textbf{Validation}} & \multicolumn{6}{c}{\textbf{Test}} \\ 
\cline{3-26}
 &  & @1 & @5 & @10 & @20 & @50 & @100 
 & @1 & @5 & @10 & @20 & @50 & @100 
 & @1 & @5 & @10 & @20 & @50 & @100 
 & @1 & @5 & @10 & @20 & @50 & @100 \\ 
\midrule
\multirow{4}{*}{\textbf{ArchivalQA}} 
 & \Vanilla 
 & 48.59 & 59.20 & 61.25 & 62.49 & 63.69 & 64.31 
 & 48.66 & 59.38 & 61.25 & 62.61 & 63.84 & 64.43 
 & 48.56 & 56.17 & 57.02 & 57.36 & 57.56 & 57.62 
 & 48.64 & 56.33 & 57.12 & 57.49 & 57.69 & 57.74 \\ 
 & \DateTag 
 & 42.06 & 52.49 & 54.71 & 56.26 & 57.62 & 58.31 
 & 43.12 & 53.27 & 55.47 & 56.96 & 58.32 & 59.04 
 & 42.03 & 49.47 & 50.38 & 50.81 & 51.03 & 51.09 
 & 43.11 & 50.32 & 51.23 & 51.63 & 51.85 & 51.92 \\ 
 & \DateToken 
 & 47.61 & 58.48 & 60.71 & 62.17 & 63.35 & 63.95 
 & 47.89 & 58.64 & 60.76 & 62.17 & 63.42 & 64.05 
 & 47.58 & 55.38 & 56.30 & 56.70 & 56.89 & 56.95 
 & 47.87 & 55.56 & 56.44 & 56.83 & 57.03 & 57.09 \\ 
 
 & \TempDPR$_{FS}$ 
 & \textbf{55.23} & \textbf{ 65.91} & \textbf{67.76} & \textbf{69.02} & \textbf{69.97} & \textbf{ 70.40} 
 & \textbf{55.29} & \textbf{66.08} & \textbf{68.02} & \textbf{64.91} & \textbf{70.14} & \textbf{70.58} 
 & \textbf{55.20} & \textbf{62.87} & \textbf{63.64} & \textbf{63.99} & \textbf{64.15} & \textbf{64.19} 
 & \textbf{55.27} & \textbf{63.00} & \textbf{63.81} & \textbf{64.11} & \textbf{64.28} & \textbf{64.32} \\

\midrule
\multirow{4}{*}{\textbf{ChroniclingAmericaQA}} 
 & \Vanilla 
 & 39.72 & 47.06 & 48.35 & 49.36 & 50.38 & 50.86 
 & 41.20 & 47.54 & 48.87 & 49.74 & 50.68 & 50.97 
 & 39.72 & 44.90 & 45.44 & 45.70 & 45.88 & 45.92 
 & 41.20 & 45.75 & 46.29 & 46.53 & 46.68 & 46.71 \\ 
 & \DateTag 
 & 40.21 & 47.00 & 48.24 & 49.16 & 50.31 & 50.82 
 & 41.20 & 47.03 & 48.41 & 49.55 & 50.39 & 50.93 
 & 40.21 & 45.00 & 45.51 & 45.76 & 45.95 & 46.00 
 & 41.20 & 45.41 & 45.98 & 46.29 & 46.43 & 46.48 \\ 
  
 & \DateToken 
 & 39.72 & 47.06 & 48.35 & 49.36 & 50.38 & 50.86 
 & 41.20 & 47.54 & 48.87 & 49.74 & 50.68 & 50.97 
 & 39.72 & 44.90 & 45.44 & 45.70 & 45.88 & 45.92 
 & 41.20 & 45.75 & 46.29 & 46.53 & 46.68 & 46.71 \\ 
 
 & \TempDPR$_{FS}$ 
 & \textbf{44.35} & \textbf{51.79} & \textbf{53.03} & \textbf{54.23} & \textbf{55.24} & \textbf{55.75} 
 & \textbf{44.83} & \textbf{51.44} & \textbf{52.85} & \textbf{53.88} & \textbf{55.01} & \textbf{55.33} 
 & \textbf{44.35} & \textbf{49.78} & \textbf{50.29} & \textbf{50.61} & \textbf{50.77} & \textbf{50.82} 
 & \textbf{44.83} & \textbf{49.52} & \textbf{50.09} & \textbf{50.38} & \textbf{50.57} & \textbf{50.60} \\ 
\bottomrule
\end{tabular}
}

\label{tab:map_ndcg-explicit_dataset}
\end{table*}

\begin{table*}[ht!]
\centering
\caption{Performance of BiTimeBERT and TS-Retriever enhanced with \TempDPR’s fusion techniques on \textit{ChroniclingAmericaQA} and \textit{ArchivalQA}. Results include Top-{1, 5, 10, 20, 50, 100} retrieval accuracy for validation and test sets, demonstrating significant improvements in temporal retrieval performance.}
\setlength\tabcolsep{5pt}
\resizebox{0.75\textwidth}{!}{
\begin{tabular}{l|l|cccccc|cccccc} 
\hline
\multirow{2}{*}{\textbf{Dataset}} & \multirow{2}{*}{\textbf{Model}} & \multicolumn{6}{c|}{\textbf{Validation Results}} & \multicolumn{6}{c}{\textbf{Test Results}} \\ 
\cline{3-14}
 &  & Top-1 & Top-5 & Top-10 & Top-20 & Top-50 & Top-100 & Top-1 & Top-5 & Top-10 & Top-20 & Top-50 & Top-100 \\ 
\midrule
\multirow{10}{*}{\textbf{ChroniclingAmericaQA}} 
 &\Vanilla + \texttt{BiTimeBERT} & 35.74 & 54.02 & 61.09   & 66.94 & 74.90 & 78.31 & 38.52 & 54.62 & 60.46 & 65.59 & 72.69 & 76.95 \\

 & \TempDPR$_{VS}+ \texttt{BiTimeBERT}$ & 37.77 & 57.19 & 65.15 & 71.41 & 78.07 & 82.45 & 38.67 & 57.14 & 63.14 &  \textbf{70.80} & 77.27 &  \textbf{82.79} \\
 & \TempDPR$_{RE} + \texttt{BiTimeBERT}$ & 35.01 & 56.13 & 63.36 & 69.37 & 77.66 &  \textbf{83.27} & 37.65 & 54.70 & 61.33 & 68.43 & 76.32 & 81.06 \\
 & \TempDPR$_{EWI}+ \texttt{BiTimeBERT}$ & 40.57 & 56.52 & 64.69 & 70.89 & 77.88 & 82.45 & 40.15 & 56.72 & 64.14 & 70.51 & 77.37 & 82.56 \\
& \TempDPR$_{FS}+\texttt{BiTimeBERT}$ & \textbf{40.86} &  \textbf{58.57} &  \textbf{65.64} &  \textbf{72.30} &  \textbf{78.47} & 82.86 &  \textbf{41.75} &  \textbf{58.96} &  \textbf{65.27} & 70.56 &  \textbf{78.69} & 81.77 \\
\cline{2-14}
 & \Vanilla + \texttt{TS-Retriever} & 47.85 & 63.93 & 69.29 & 73.60 & 79.53 & 83.43 & 46.65 & 62.35 & 69.06 & 74.27 & 78.77 & 82.56 \\

 & \TempDPR$_{VS}+\texttt{TS-Retriever}$ & 49.96 & 67.18 & 72.30 & 77.58 & 82.13 & 85.87 & 50.99 & 65.82 & 72.06 &  \textbf{77.43} &  \textbf{82.08} & 84.93 \\
 & \TempDPR$_{RE}+\texttt{TS-Retriever}$ & 51.22 & 65.51 & 72.14 & 76.01 & 81.77 & 84.93 & 52.80 & 67.26 & 72.54 & 77.17 & 82.70 &  \textbf{87.00} \\
 & \TempDPR$_{EWI}+\texttt{TS-Retriever}$ & 49.15 & 64.34 & 71.57 & 76.77 & 82.05 & 85.78 & 49.96 & 65.27 & 70.01 & 75.30 & 80.74 & 83.98 \\
  & \TempDPR$_{FS}+\texttt{TS-Retriever}$ &  \textbf{54.02} &  \textbf{67.99} &  \textbf{73.19} &  \textbf{77.82} &  \textbf{83.35} &  \textbf{86.11} &  \textbf{53.20} &  \textbf{67.40} &  \textbf{73.09} &  \textbf{77.43} & 82.00 & 86.03 \\
\midrule
\multirow{10}{*}{\textbf{ArchivalQA}} 
 & \Vanilla + \texttt{BiTimeBERT} & 61.71 & 81.20 & 86.56 & 89.99 & 93.88 & 95.52 & 60.22 & 81.20 & 86.70 & 90.48 & 94.09 & 95.62 \\

 & \TempDPR$_{VS}+\texttt{BiTimeBERT}$ & 63.70 & 83.09 & 88.48 & 92.12 & 95.14 & 96.63 & 63.88 & 83.14 & 88.33 & 91.96 & 95.00 & 96.49 \\
 & \TempDPR$_{RE}+\texttt{BiTimeBERT}$ & 60.01 & 80.60 & 85.84 & 90.00 & 93.79 & 95.89 & 61.93 & 80.96 & 85.93 & 90.20 & 93.76 & 95.72 \\
 & \TempDPR$_{EWI}+\texttt{BiTimeBERT}$ & 61.28 & 80.93 & 86.33 & 90.19 & 93.46 & 95.50 & 60.52 & 81.22 & 86.57 & 90.76 & 94.30 & 95.90 \\
  & \TempDPR$_{FS}+\texttt{BiTimeBERT}$ &  \textbf{64.93} &  \textbf{84.36} &  \textbf{89.04} &  \textbf{92.51} &  \textbf{95.63} &  \textbf{96.95} &  \textbf{65.66} &  \textbf{84.63} &  \textbf{89.64} &  \textbf{92.55} &  \textbf{95.49} &  \textbf{96.81} \\
\cline{2-14}
 & \Vanilla + \texttt{TS-Retriever} & 63.44 & 83.74 & 88.70 & 92.53 & 95.88 & 97.33 & 64.68 & 84.75 & 89.44 & 92.92 & 95.76 & 97.29 \\

 & \TempDPR$_{VS}+\texttt{TS-Retriever}$ & 69.20 & 87.82 & 92.14 & 94.77 & 97.03 &  \textbf{98.15} & 69.88 & 88.09 & 92.61 & 95.31 & 97.62 & 98.50 \\
 & \TempDPR$_{RE}+\texttt{TS-Retriever}$ & 70.11 & 88.05 & 92.17 & 94.66 & 96.93 & 98.06 & 70.65 & 88.42 & 92.38 & 95.10 & 97.39 & 98.37 \\
 & \TempDPR$_{EWI}+\texttt{TS-Retriever}$ &  \textbf{70.56} & 87.46 & 91.62 & 94.40 & 96.74 & 97.86 &  \textbf{70.96} & 88.06 & 92.07 & 94.94 & 97.08 & 98.18 \\
  & \TempDPR$_{FS}+\texttt{TS-Retriever}$ & 69.94 &  \textbf{88.20} &  \textbf{92.23} &  \textbf{95.26} &  \textbf{97.12} &  \textbf{98.15} & 70.37 &  \textbf{88.82} &  \textbf{92.69} &  \textbf{95.41} &  \textbf{97.52} &  \textbf{98.45} \\
\bottomrule
\end{tabular}
}
\label{tab:combined_results2}
\end{table*}
\subsection{Selection of Negative Passages}\label{sec:negativepassages}

Commonly, in retrieval tasks, positive examples are often available, whereas negative examples must be selected from a large and diverse pool of irrelevant passages~\cite{karpukhin2020dense}. 
Careful selection of appropriate negative examples rather than random choice is a critical aspect of learning a high-quality encoder~\cite{izacard2021unsupervised,karpukhin2020dense}, however, this tends to be overlooked in current retrieval setups. 

We adopt the following strategies for testing the selection of negative passages: (1) \textbf{Random Negative}: We randomly select passages from the corpus that do not contain the correct answer, ensuring by this that they are irrelevant to the query. (2) \textbf{Temporal Negative (Same Year)}: We randomly select passages from the same year as the positive passage's date that do not contain the answer. This helps evaluate the model's ability to differentiate temporally aligned passages based on their content relevance. (3) \textbf{Temporal Negative (Different Year)}: We randomly select passages from a year different than the year of a positive passage, ensuring also that they do not contain the correct answer. This approach helps assess the impact of temporal misalignment on retrieval performance. 
\subsection{Settings}
The \TempDPR model was trained using an in-batch negative sampling approach with a batch size of 32. For our experiments, we trained the \TempDPR for $5$ epochs with a learning rate set to $1 \times 10^{-5}$, using the Adam optimizer and a warm-up ratio of $0.1$. All experiments (except Section \ref{sec:Negative}) were conducted with 4 randomly selected negative passages per query, ensuring that none of the negative passages contained the correct answer.

To evaluate the trained retrievers, we employed standard metrics, including Top-\(k\) accuracy~\cite{karpukhin2020dense}, normalized discounted cumulative gain (nDCG), and mean average precision (MAP).

\vspace{-2mm}

\section{Experimental Results}
\label{sec:experiments}



\subsection{Impact of Temporal Fusion Strategies}
We first start with investigating the performance of different temporal fusion techniques in \TempDPR model, such as Feature Stacking (FS), Vector Summation (VS), Relative Embeddings (RE), and Element-Wise Interaction (EWI) with state-of-the-art temporal retrieval models, BiTimeBert~\cite{wang2023bitimebert} and TS-Retriever~\cite{wu2024time}, as well as the baseline \Vanilla \cite{karpukhin2020dense}. The results in Table~\ref{tab:combined_results} highlight the effectiveness of our temporal fusion techniques across both the \textit{ChroniclingAmericaQA} and \textit{ArchivalQA} datasets. On the \textit{ChroniclingAmericaQA} dataset, \TempDPR$_{FS}$ outperforms all other models, achieving a Top-1 accuracy of 50.22\% on the validation set. This represents a substantial improvement over BiTimeBert, which achieves 35.74\%, and TS-Retriever, which reaches 47.85\%.  In terms of Top-100 accuracy, \TempDPR$_{FS}$ achieves 86.22\%, surpassing TS-Retriever and BiTimeBert by 2.79 and 7.91\%, respectively. Similarly, on the \textit{ArchivalQA} dataset, \TempDPR$_{FS}$ achieves state-of-the-art results, with a Top-1 accuracy of 69.72\% on the validation set. This is a clear improvement over BiTimeBert (61.71\%) and TS-Retriever (63.44\%). For higher $k$ values, such as the Top-100 accuracy, \TempDPR$_{FS}$ reaches 97.77\%, representing a notable improvement over both TS-Retriever (97.33\%) and BiTimeBert (95.52\%). These results demonstrate the capability of \TempDPR to capture and align temporal signals more effectively than competing methods.

\subsection{Retrieval Performance Analysis}
\label{sec:main_result}

In this section, we present a comprehensive comparison of \TempDPR with baseline methods, focusing on its overall retrieval performance across various metrics, such as top-$k$ accuracy, nDCG, and MAP. We evaluate \TempDPR alongside \texttt{\Vanilla}, \texttt{\DateTag}, and \texttt{\DateToken} models.

Table~\ref{tab:top-explicit_dataset} shows the retrieval performance of \TempDPR against three baselines on the \textit{ArchivalQA} and \textit{ChroniclingAmericaQA} datasets, using top-$k$ accuracy ($k \in \{1, 5, 10, 20, 50, 100\}$). Across both datasets, \TempDPR outperforms the \texttt{\Vanilla}, \texttt{\DateTag}, and \texttt{\DateToken} models, demonstrating the benefit of directly incorporating temporal embeddings into the retrieval process. On \textit{ChroniclingAmericaQA}, \TempDPR achieves a top-100 accuracy of 86.22\% on the validation set, compared to 83.53\% for \texttt{\Vanilla}. Table~\ref{tab:map_ndcg-explicit_dataset} summarises the nDCG and MAP metrics.
\TempDPR achieves an nDCG@10 of 53.03 on the validation set of \textit{ChroniclingAmericaQA}, outperforming the \texttt{\DateTag} and \texttt{\DateToken} models by over $4\%$. MAP shows a similar trend, with \TempDPR achieving a MAP@20 of 53.88\% on the test set, significantly higher than the 49.74\% obtained by the \Vanilla baseline. 

Similar trends are observed on \textit{ArchivalQA}, where \TempDPR demonstrates superior performance in ranking. As shown in Table~\ref{tab:map_ndcg-explicit_dataset}, \TempDPR achieves nDCG@10 score of 67.76 on the validation set, surpassing the \texttt{\DateTag} and \texttt{\DateToken} approaches, which reach 54.71 and 60.71, respectively.  Also, \TempDPR outperforms the other methods significantly, achieving a MAP@10 of 63.64 on the validation set. The \texttt{\DateTag} and \texttt{\DateToken} models achieve MAP of 50.38 and 56.30, respectively. These results highlight that \TempDPR not only retrieves semantically relevant passages but also ranks them more accurately according to their temporal relevance, resulting in improvement compared to baseline methods.

To validate \TempDPR’s performance improvements, we conducted paired t-tests comparing Top-1, Top-50, and Top-100 accuracies with baselines (\texttt{\Vanilla}, \texttt{\DateTag}, \texttt{\DateToken}). Table~\ref{tab:statistical_analysis} shows the p-values for these comparisons, all significantly below 0.05, confirming the statistical significance of \TempDPR’s results. For instance, the p-values for Top-1 accuracy were $1.39 \times 10^{-5}$, $4.18 \times 10^{-7}$, and $2.13 \times 10^{-5}$ for \texttt{\Vanilla}, \texttt{\DateTag}, and \texttt{\DateToken}, respectively. Similar trends were observed for Top-50 and Top-100 metrics, demonstrating the robustness of \TempDPR’s improvements across retrieval metrics.


\begin{table}[h!]
\centering
\caption{Paired t-test results for Top-$k$ accuracy comparisons between \TempDPR and baselines (\texttt{\Vanilla}, \texttt{\DateTag}, and \texttt{\DateToken}).}
\label{tab:statistical_analysis}
\resizebox{0.40\textwidth}{!}{ 
\begin{tabular}{lccc}
\toprule
\textbf{Metric} & \textbf{\Vanilla DPR} & \textbf{\DateTag} & \textbf{\DateToken} \\
\midrule
Top-1 Accuracy  & $1.39 \times 10^{-5}$ & $4.18 \times 10^{-7}$ & $2.13 \times 10^{-5}$ \\
Top-50 Accuracy & $2.57 \times 10^{-4}$ & $3.42 \times 10^{-6}$ & $1.89 \times 10^{-4}$ \\
Top-100 Accuracy & $7.11 \times 10^{-5}$ & $1.26 \times 10^{-6}$ & $9.84 \times 10^{-5}$ \\
\bottomrule
\end{tabular}}
\end{table}

\subsection{Enhancing Temporal Models with Fusion Strategies}

To demonstrate the flexibility and impact of \TempDPR’s fusion techniques, we evaluate their ability to enhance the performance of existing temporal retrieval models, including BiTimeBERT~\cite{wang2023bitimebert} and TS-Retriever~\cite{wu2024time}. By integrating temporal embeddings and leveraging fusion strategies, we showcase how \TempDPR can amplify the strengths of SOTA temporal models.

Table~\ref{tab:combined_results2} demonstrates how incorporating \TempDPR’s fusion techniques enhances the performance of existing temporal retrieval models, BiTimeBERT~\cite{wang2023bitimebert} and TS-Retriever~\cite{wu2024time}, across the \textit{ChroniclingAmericaQA} and \textit{ArchivalQA} datasets. 
On the \textit{ChroniclingAmericaQA} dataset, BiTimeBERT alone achieves a Top-1 accuracy of 35.74\% on the validation set. However, when enhanced with \TempDPR’s fusion techniques, its performance improves substantially. For example, with feature stacking (\TempDPR$_{FS}+\texttt{BiTimeBERT}$), the Top-1 accuracy increases to 40.86\%, reflecting an improvement of over 5\%. Even though TS-Retriever incorporates temporal constraints through tailored negative sampling, it also benefits significantly from \TempDPR’s fusion techniques. On the same dataset, TS-Retriever achieves a Top-1 accuracy of 47.85\% on the validation set. By integrating \TempDPR with feature stacking ( \TempDPR$_{FS}+\texttt{TS-Retriever}$), the Top-1 accuracy improves to 54.02\%. This demonstrates how fusion techniques can amplify the strengths of TS-Retriever, aligning temporal signals more effectively with semantic content. For instance, the Top-100 accuracy increases from 83.43\% to 86.11\%, highlighting the robustness of feature stacking in retaining temporal relevance across diverse query-document pairs. 

Similar trends are observed on the \textit{ArchivalQA} dataset. BiTimeBERT alone achieves a Top-1 accuracy of 61.71\%, but this increases to 64.93\% with \TempDPR$_{FS}+\texttt{BiTimeBERT}$. TS-Retriever shows an even more pronounced improvement, with its Top-1 accuracy rising from 63.44\% to 69.94\% when combined with \TempDPR and feature stacking.

\subsection{Negative Document Selection Strategies}
\label{sec:Negative}
In this section, we analyze the impact of different negative sampling strategies on retrieval performance, focusing on three approaches: Random sampling, Same-Year sampling, and Different-Year sampling. We test the above strategies involving also the variation in the number of negative documents used during training to explore how these choices affect retrieval accuracy in the context of temporally sensitive datasets like \textit{ChroniclingAmericaQA} and \textit{ArchivalQA}.

\begin{figure*}
    \centering
    \includegraphics[width=0.65\textwidth]{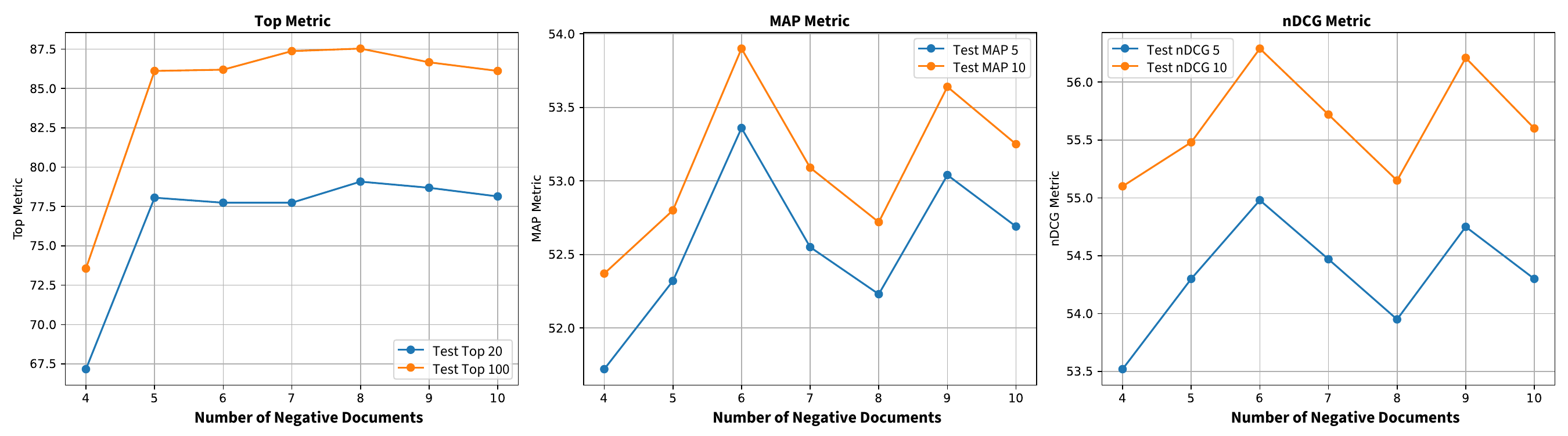}
    \caption{Top-k accuracy, MAP, and nDCG Metrics with the different number of negative documents included during training for \textit{ChroniclingAmericaQA}.}
    \label{fig:number-neggitive-examples}
\end{figure*}

Table \ref{tab:ablation} shows the evaluation results of these strategies for both validation and test sets of \textit{ChroniclingAmericaQA} and \textit{ArchivalQA} datasets, using metrics nDCG and MAP. For instance, in \textit{ChroniclingAmericaQA}, the Different-Year strategy achieves an nDCG@10 score of 55.04 with 4 negative documents, compared with the Random and Same-Year approaches, which reach 53.03 and 53.51, respectively. In \textit{ArchivalQA}, the Same-Year strategy with 5 negative documents achieves the highest nDCG and MAP scores among all the strategies, with nDCG@10 reaching 66.73 on the validation set. Hence, the choice between Same-Year and Different-Year seems to depend on a dataset used. However, on average, either of them seems to be a better solution than the random selection strategy.

\begin{table}[ht]

\setlength\tabcolsep{5pt}
\centering
\caption{Evaluation of Different Negative sampling Modes on \textit{ArchivalQA} and \textit{ChroniclingAmericaQA} Datasets. The table shows Top-k, nDCG, and MAP metrics for both Validation and Test sets.}
\resizebox{0.45\textwidth}{!}{  
\begin{tabular}{l|l|c|cc|cc|cc|cc}
\hline
\multirow{2}{*}{\textbf{Dataset}} & \multirow{2}{*}{\textbf{Mode}} & \multirow{2}{*}{\textbf{\# Neg}}  & \multicolumn{4}{c|}{\textbf{Val}} & \multicolumn{4}{c}{\textbf{Test}}  \\
 & &  & \multicolumn{2}{c|}{\textbf{nDCG}}& \multicolumn{2}{c|}{\textbf{MAP}}   & \multicolumn{2}{c|}{\textbf{nDCG}}  & \multicolumn{2}{c}{\textbf{MAP}} \\
 &  & \textbf{Documents}  & 5 & 10 &  5 & 10   & 5 & 10  & 5 & 10  \\
\hline
\multirow{6}{*}{\textbf{ArchivalQA}} 
& Random & 4   &60.13  &62.20    &  57.10 & 57.90    &  60.34&  62.20 &   57.50 & 58.26 \\

& Same Year & 4   & 63.39 & 65.47 & 60.33 & 61.20  &  63.12 & 65.15 &  60.03 &  60.88 \\

& Diff. Year & 4   &62.69 & 64.81   &  59.68  & 60.56   &62.78& 64.88 &   59.61 &  60.49\\

& Random & 5   &61.25  &63.22   &  58.12 & 58.93   &  61.49&  63.53  &  58.41 & 59.25  \\

& Same Year & 5  & \textbf{64.92} & \textbf{66.73}   &  \textbf{61.86} & \textbf{62.60} &    \textbf{64.79} & \textbf{66.71}  & \textbf{61.75} &\textbf{62.55}  \\

& Diff. Year & 5     & 64.05 & 65.96 &  60.93 & 61.73  &64.44 &66.38 &61.51 & 62.31   \\

\hline
\multirow{6}{*}{\textbf{ChroniclingAmericaQA}} 
& Random & 4  & 51.79 & 53.03  & 49.78 & 50.29 &    51.44 & 52.82  & 49.50 &  50.09\\

& Same Year & 4  & 52.19 & 53.51  & 50.16 & 50.71   & 51.69 & 52.84  & 49.48 & 49.95\\

& Diff. Year & 4   & 53.99 & 55.04 &  52.01 & 52.46  & 53.52 & 55.10  & 51.72& 52.37\\

& Random & 5   &  52.73 &  54.31   & 51.00 & 51.65  & 51.38 &  53.13   & 49.66 & 50.40 \\

& Same Year & 5  &52.20 & 53.35  & 50.40 & 50.88  &  51.75 & 53.11   &  49.99 &50.56  \\

& Diff. Year & 5   & \textbf{54.07} & \textbf{55.56}  & \textbf{52.43} & \textbf{53.05}  & \textbf{54.30} & \textbf{55.48}  & \textbf{52.32} & \textbf{52.80} \\

\hline

\end{tabular}}

\label{tab:ablation}
\end{table}

Figure \ref{fig:number-neggitive-examples} shows the relationship between the number of negative documents and the retrieval performance on the \textit{ChroniclingAmericaQA} dataset for the Different-Year method, measured across various metrics. As the number of negative samples increases from 4 to 10, there is a noticeable improvement in both Top-20 and Top-100 accuracy, as well as in MAP and nDCG. The Top-20 and Top-100 curves show a relatively steady increase, suggesting that the model benefits from more challenging negative examples, which enhances its ability to differentiate between relevant and irrelevant passages. Similarly, the MAP and nDCG metrics, which assess ranking quality, show improvements, particularly for the top 5 and 10 retrieved passages. 

\section{Implicit Temporal Questions}\label{sec: implicit}

Implicit temporal questions, where queries lack explicit temporal references, present significant challenges for retrieval models. To address this, we developed a \textit{query date prediction model}. The model is based on BERT and is trained on the \textit{Event Sentence} dataset~\cite{wallat2024temporal}, which contains short event descriptions annotated with their corresponding years. The dataset includes a total of 22,399 instances, split into 17,919 for training, 2,240 for validation, and 2,240 for testing, spanning 21 classes representing years from 1987 to 2007.

The trained model achieves a \textbf{Mean Absolute Error (MAE)} of \textbf{3.51} years, a \textbf{Mean Squared Error (MSE)} of \textbf{25.65} years, and an overall \textbf{accuracy} of \textbf{20\%} for predicting the correct year of the event described in the target short text. Despite the inherent difficulty of this task, the query date prediction model provides a meaningful temporal context that can be effectively integrated into retrieval models for improved performance of implicit temporal questions.

\begin{table*}[ht]
\centering
\caption{ Results for Top-\{1,5,10,20,50,100\} for the implicit questions from \textit{ArchivalQA} datasets, across validation and test sets.}
\setlength\tabcolsep{5pt}
\resizebox{0.65\textwidth}{!}{  
\begin{tabular}{l|l|cccccc|cccccc} 
\hline
\multirow{2}{*}{\textbf{Dataset}} & \multirow{2}{*}{\textbf{Mode}} &  \multicolumn{6}{c|}{\textbf{Validation Results}} & \multicolumn{6}{c}{\textbf{Test Results}} \\ 
\cline{3-14}

\cline{3-14}
 &  & Top-1 & Top-5 & Top-10 & Top-20 & Top-50 & Top-100 & Top-1 & Top-5 & Top-10 & Top-20 & Top-50 & Top-100 \\ 
\midrule
\multirow{8}{*}{\textbf{ArchivalQA}} 
 & \Vanilla & 43.10 & 66.10  & 73.60  & 79.50 & 85.50  & 89.50 & 41.90 & 65.20 & 72.60  & 78.60 & 85.80 & 89.70 \\

 & BiTimeBERT &27.70 & 50.70 &59.80  & 68.00 & 75.50  & 81.80  &25.90 & 49.30 & 57.70  & 65.20 & 75.10 &81.90 \\ 
 
 & TS Retriever & 39.60 & 65.90 & 74.60 & 80.40 & 86.70 & 90.90  & 40.30 &65.50  & 73.30 &\textbf{80.50}  &87.90 & 92.00\\ 

 & \DateTag & 42.00 & 64.60 & 71.90 & 77.60 & 84.30 & 88.10 & 39.10 & 62.10 & 70.90 & 77.10 & 84.30 & 88.20\\ 
 
 & \DateToken & 42.60 & 65.90 & 72.90 & 79.00 & 85.10 & 89.00 & 41.40 & 64.70 & 72.50 & 78.50 & 84.40 & 88.90 \\ 

 & \TempDPR$_{VS}$ &\textbf{45.00} & 67.30 & \textbf{74.70} & \textbf{80.70} & 86.40 & 90.10 & 41.70 &  66.40 & 73.80 & 80.00 & 85.10 & 89.60 \\ 
  
 & \TempDPR$_{EWI}$&43.50  & 65.60 & 72.40 & 77.70 & 84.80 &  88.50 & 40.20 & 63.70 &  70.40 & 76.60 & 83.80 &88.00 \\ 
  
 & \TempDPR$_{FS}$& 44.80 &  \textbf{67.40} & 74.10&79.60 & \textbf{86.50} & \textbf{90.20} &\textbf{43.30} & \textbf{66.40} &\textbf{74.50 } & 79.60 & \textbf{86.10} & \textbf{89.70}\\


\midrule

\end{tabular}
}

\label{tab:top-implicit_dataset}
\end{table*}

\begin{figure}[h!]
    \centering
    \includegraphics[width=0.45\textwidth]{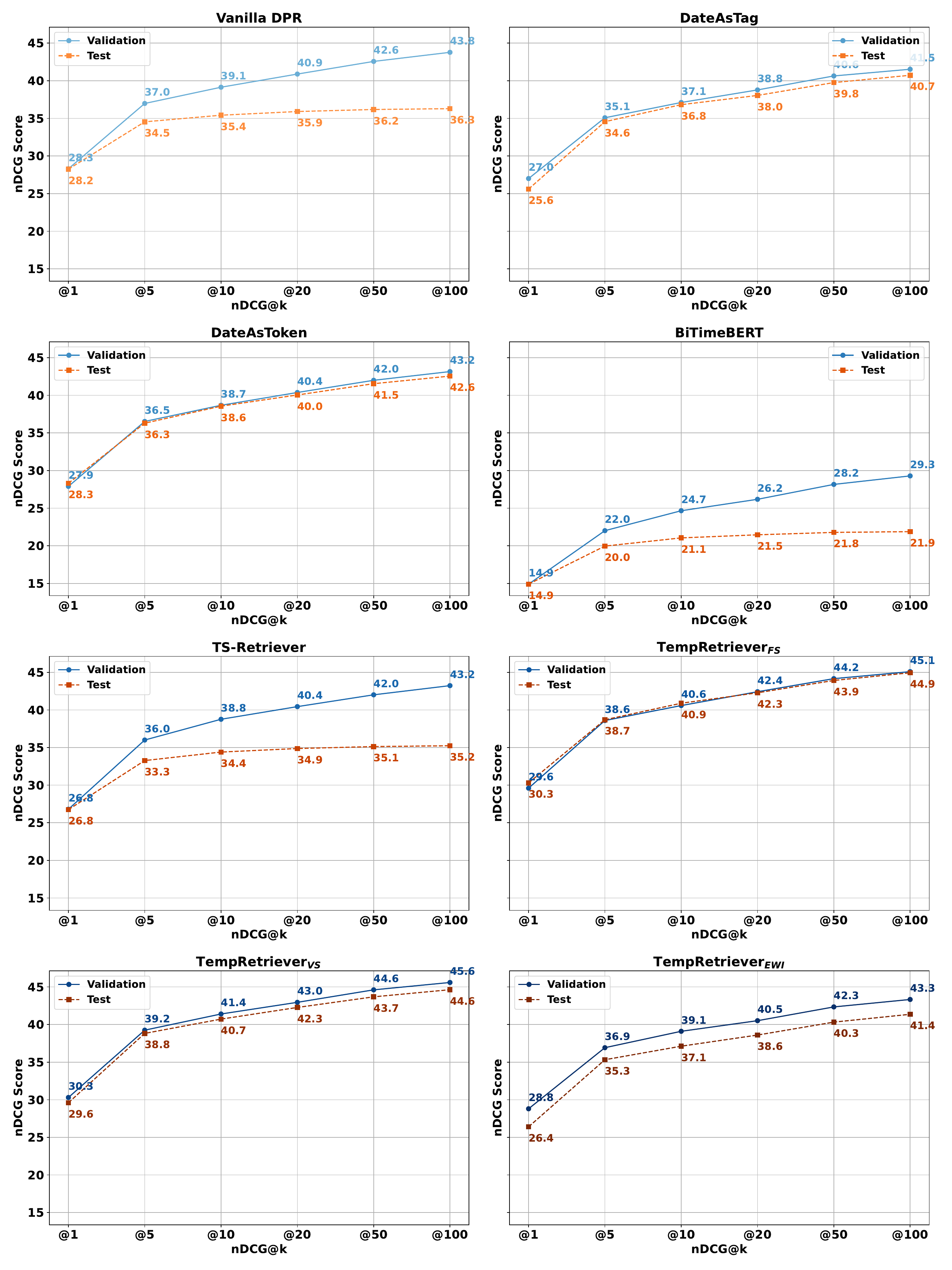} 
    \caption{Comparison of nDCG  for implicit questions across different retrieval models on the ArchivalQA dataset. Each subplot represents a specific retrieval model, showing performance for validation and test datasets at varying cut-off points (\texttt{nDCG@k}).}

    \label{fig:implicit_ndcg_results}
\end{figure}
\vspace{-2mm}
\subsection{Implicit Temporal Questions in Retrieval}
\label{implict_question}

The evaluation on the \textit{ArchivalQA} dataset demonstrates the effectiveness of the proposed \TempDPR framework in addressing implicit temporal questions, as shown in Table~\ref{tab:top-implicit_dataset} and Figure~\ref{fig:implicit_ndcg_results}. \TempDPR consistently outperforms baseline models and pretrained temporal models across validation and test sets. For example, \Vanilla achieves a Top-1 accuracy of 43.10\% and a Top-100 accuracy of 89.50\% on the validation set, while \DateToken slightly improves with a Top-100 accuracy of 89.00\%. Among temporal models, BiTimeBERT performs the worst (Top-1: 27.70\%, Top-100: 81.80\%), highlighting its limitations in reasoning about implicit temporal information. TS-Retriever performs better, reaching a Top-1 accuracy of 39.60\% and a Top-100 accuracy of 90.90\%.

\TempDPR models, particularly \TempDPR$_{FS}$, achieve state-of-the-art results, with a Top-1 accuracy of 44.80\% and a Top-100 accuracy of 90.20\%. \TempDPR$_{VS}$ achieves the highest Top-1 accuracy of 45.00\%. Ranking quality metrics (Figure~\ref{fig:implicit_ndcg_results}) further demonstrate \TempDPR’s ability to prioritize relevant passages. For instance, at \texttt{nDCG@1}, \Vanilla achieves 28.30\%, which improves to 29.60\% and 30.30\% with \TempDPR$_{FS}$ and \TempDPR$_{VS}$, respectively. This trend persists across all \(k\)-values, with \TempDPR$_{FS}$ achieving the highest scores, such as 45.09\% at \texttt{nDCG@100}, compared to 43.78\% for \Vanilla and 29.29\% for BiTimeBERT.

\subsection{Addressing Different Question Types}

\begin{figure}[ht!]
\includegraphics[width=0.4\textwidth]{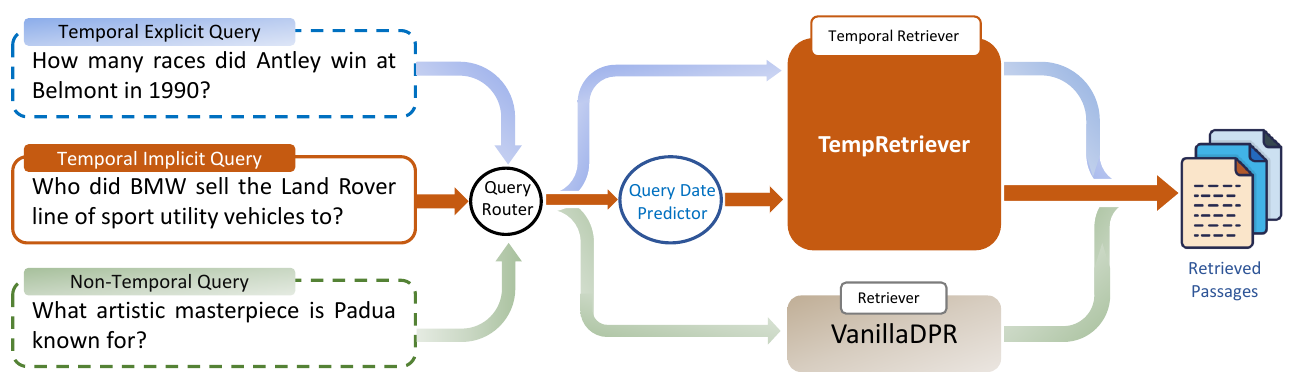}
\caption{
A simple unified framework for handling different questions types
using a Query Router.
}
\label{fig:query_routing}
\end{figure}

We next explain how to construct a unified framework for handling different question types. Temporal retrieval systems must address diverse question types, including explicit temporal, implicit temporal, and non-temporal queries. As illustrated in Fig.~\ref{fig:query_routing}, a query routing mechanism can be implemented to dynamically classify and assign questions to the appropriate retriever. Explicit temporal questions, which contain direct references to dates, are routed to \TempDPR, leveraging its ability to integrate temporal embeddings and fusion techniques. 

Implicit temporal questions that lack explicit date mentions are routed to the query date prediction model, which predicts a temporal context based on the query’s content. The predicted date is then used to enhance the retrieval process through \TempDPR, aligning the inferred temporal context with the document timestamps. Finally, non-temporal queries, which do not involve temporal elements are directed to a general retriever like \Vanilla. This simple routing framework optimizes retrieval performance across diverse query types by aligning each query with the most effective retriever.

\begin{table}[!ht]
\centering
\caption{Zero-shot results of Temporal QA on \textit{ChroniclingAmericaQA} and \textit{ArchivalQA} measured by Exact Match, Recall and Contain. In the open-book setting, we include the top-1 retrieved document. Bold fonts indicate best results for a model, while we underline the best results across models.
}
\resizebox{0.45\textwidth}{!}{  
\setlength\tabcolsep{3pt}
\begin{tabular}{@{}l |l| ccc |ccc | ccc | ccc @{}}
\toprule
 & & \multicolumn{6}{c|}{\textbf{Val}} & \multicolumn{6}{c}{\textbf{Test}}\\

& \multirow{3}{*}{\textbf{Retriever} }&  \multicolumn{3}{c|}{\textbf{ChroniclingAmericaQA}} & \multicolumn{3}{c|}{\textbf{ArchivalQA}} & \multicolumn{3}{c|}{\textbf{ChroniclingAmericaQA}} & \multicolumn{3}{c}{\textbf{ArchivalQA}} \\

    &                 & EM & Recall  & Con  & EM & Recall  & Con & EM & Recall  & Con & EM & Recall  & Con \\
\midrule

 \multirow{4}{*}{\textbf{Gemma-2-2B}~\cite{team2024gemma}} & \Vanilla  &  14.13  &  25.66 & 16.97 & 38.10 & 57.49 & 44.13 &15.78 &26.68 & 18.23 & 35.78 & 57.16 & 44.71\\

&\DateTag &  13.64    & 25.38 & 17.14 & 35.56 & 53.58 & 40.63 & 15.70 & 26.73 & 18.31 & 33.31  & 53.52  & 41.62 \\

&\DateToken     & 14.21 &  25.71 &  17.38 & 38.12 & 57.49& 44.20& 15.54 & 26.85 &18.15  & 35.86 & 56.84 &  44.31\\

&\TempDPR   & \textbf{15.10} & \textbf{27.16} & \textbf{18.92} & \textbf{38.32} & \textbf{58.30} & \textbf{44.67} & \textbf{16.54} & \textbf{27.29} & \textbf{19.07} & \textbf{36.60} & \textbf{57.91} & \textbf{\underline{45.32}}\\

\midrule
 \multirow{4}{*}{\textbf{Llama-2-7B }~\cite{touvron2023llama}} &\Vanilla  &  14.37   &  24.65 & 15.75 & 38.79 & 55.48 & 42.49 &15.78 &24.85 & 15.86 & 36.57 & 55.40 & 43.00\\

&\DateTag &  13.24  & 23.50 & 16.57 & 36.66 & 52.21 & 39.71 & 15.07 &24.72 &16.65 & 34.27 &   51.54 & 39.68 \\

&\DateToken&  14.05 & 24.33 & 16.08 &  39.07 & 56.06 & 42.77 & 17.04 & 26.36 &17.99 & 36.77 &  55.24 & 42.59 \\

&\TempDPR    & \textbf{14.45} & \textbf{25.98} & \textbf{18.03} & \textbf{39.94} &\textbf{57.38} &\textbf{43.71}  & \textbf{17.73} & \textbf{27.48} &\textbf{18.91} & \textbf{37.51} & \textbf{56.43} &\textbf{43.56} \\

\midrule
 \multirow{4}{*}{\textbf{Llama-3-8B}~\cite{llama3modelcard}} &\Vanilla  & 14.86  & 26.01 &17.30  & 40.87 &57.56 & 44.52 & 17.28 &  27.15 & 18.31 & 38.23 & 57.17 & 44.59\\

&\DateTag & 16.08 & 26.43 & 18.27 & 37.25 & 53.16 & 40.68 & 16.17 & 26.76 & 18.07 & 35.28 &  53.08 & 41.20 \\

&\DateToken& 16.57  & 27.33 & 18.03 & 40.97 & 57.57 & 44.43 & 17.99 & 29.19 &19.88 & 38.28 & 57.00 & 44.31\\

&\TempDPR& \textbf{\underline{16.90}} &  \textbf{\underline{28.52}} &\textbf{\underline{19.65}}  & \textbf{\underline{41.92}} & \textbf{\underline{58.52}} & \textbf{\underline{45.21}} & \textbf{\underline{18.36}} &  \textbf{\underline{30.74}} & \textbf{\underline{20.78}} & \textbf{\underline{39.25}} & \textbf{\underline{58.14}} & \textbf{45.14}\\

\bottomrule

\end{tabular}
}

\label{tab:rag_results}
\end{table}

\section{RAG for Temporal Question Answering}
The integration of \TempDPR into the Retrieval-Augmented Generation (RAG) system involves replacing the traditional retriever component with \TempDPR, which retrieves the top-k passages most relevant to a given query by considering both semantic and temporal relevance. These passages are then passed to a pre-trained large language model (LLM), such as Llama-3-8B~\cite{llama3modelcard}, for answer generation. The pipeline includes three main steps: encoding the query with its temporal context to create dense representations, ranking passages based on their combined semantic and temporal alignment, and generating answers using the retrieved passages as input for the LLM.

For the experiments, Llama-3-8B was used as the LLM, and its performance was evaluated on the \textit{ChroniclingAmericaQA} and \textit{ArchivalQA} datasets. The retriever retrieved top-1 passages for each query, which were fed into the LLM for supporting answer generation. Metrics such as Exact Match (EM), Recall (R), and Contains (Con) were used to compare the performance of RAG systems utilizing \TempDPR, \Vanilla, and other baseline models.

Table~\ref{tab:rag_results} presents the results of the RAG framework for temporal question answering on explicit datasets. \TempDPR consistently outperformed other retrieval methods across all evaluated models and datasets. For example, when paired with Llama-3-8B, \TempDPR achieved an EM score of 16.90\% and a recall of 28.52\% on the validation set of \textit{ChroniclingAmericaQA}. These scores were significantly higher than those obtained by the \DateTag and \DateToken models, which, while showing improvements over \Vanilla, fell short of \TempDPR’s performance. The \DateToken model achieved an EM score of 16.57\% on the same dataset, slightly below \TempDPR.

Across different LLMs, \TempDPR provided superior inputs for the generation task, as reflected in its higher metrics. For instance, with Llama-3-8B, \TempDPR achieved an EM score of 19.65\% on \textit{ChroniclingAmericaQA} and 45.21\% on \textit{ArchivalQA}, demonstrating its robustness in handling temporally-sensitive queries. These results underline \TempDPR's ability to deliver temporally-aligned and contextually-relevant passages, enabling more accurate and effective answer generation in RAG systems.

\vspace{-2mm}
\section{Conclusion}
In this paper, we introduced \TempDPR, a novel framework for Temporal Information Retrieval that integrates temporal information into dense passage retrieval using fusion techniques. By incorporating temporal signals directly into query and document representations, \TempDPR bridges the gap between semantic and temporal alignment, achieving superior performance across multiple retrieval metrics. Experiments on ArchivalQA and ChroniclingAmericaQA datasets demonstrate that \TempDPR significantly outperforms state-of-the-art baselines, including \Vanilla, \DateToken, and \DateTag models, as well as temporal retrieval models like BiTimeBERT and TS-Retriever.

We also evaluated \TempDPR in a Retrieval-Augmented Generation (RAG) setup, where it provided high-quality inputs to large language models, leading to improved answer generation in temporal question answering. Looking forward, we plan to enhance \TempDPR's handling of implicit temporal queries by integrating more advanced query date prediction models.  



\bibliographystyle{ACM-Reference-Format}
\bibliography{sample-base}

\appendix

\end{document}